\documentstyle[aps]{revtex}

\input{tcilatex}

\begin{document}
\title{A Global Algorithm for Training Multilayer Neural Networks}
\author{Hong Zhao\thanks{%
To whow the correspondings should be addressed. Email: zhaoh@xmu.edu.cn.}
and Tao Jin}
\address{Physics Department of Xiamen University, \\
and the Institute of Theoretical Physics and Astrophysics of Xiamen\\
University, \\
Xiamen 361005, China}
\maketitle

\begin{abstract}
We present a global algorithm for training multilayer neural networks in
this Letter. The algorithm is focused on controlling the local fields of
neurons induced by the input of samples by random adaptations of the
synaptic weights. Unlike the backpropagation algorithm, the networks may
have discrete-state weights, and may apply either differentiable or
nondifferentiable neural transfer functions. A two-layer network is trained
as an example to separate a linearly inseparable set of samples into two
categories, and its powerful generalization capacity is emphasized. The
extension to more general cases is straightforward.
\end{abstract}

The multilayer neural network, trained by the backpropagation(BP) algorithm,
is currently the most widely used neural network since it can solve linearly
inseparable classification problems\cite{Hagan,Rumelhart}. The BP algorithm
is responsible for the rebirth of neural networks.

However, the BP algorithm has several limitations. Firstly, it requires the
neural transfer functions to be differentiable in order to calculate the
derivatives with respect to the synaptic weights. Secondly, the performance
index to be minimized is constantly to be mean square error, because a
non-quadratic performance index may result in very complex performance
surface. Finally, the synaptic weights obtained by this algorithm are
continuous as a consequence of its updating equation of synaptic weights.
These limitations are also inherent in variations of the BP algorithm. The
last is indeed also a limitation for most of the learning rules for training
single-layer neural networks. Discrete synaptic states have not only the
advantage for digital hardware realization but also an experimental reality.
Recent experiments have shown that the synaptic states in certain real
neural network systems may be discrete \cite{Petersen,Connor,Abarbanel}.

The BP algorithm is a local learning rule. Most of the learning rules for
training single-layer neural networks, such as the perceptron rule, the Hebb
rule, and the Widrow-Hoff rule, are local rules. When training a network
using a local rule, one inputs the samples into the network one by one, and
each time the synaptic weights are updated independently on other samples. A
step of update of synaptic weights induced by the input of a sample is an
optimal solution for this sample, but not for other samples.

In principle, it is more favorable if each step of the update is an optimal
solution for all the samples. This requires the consideriation of the whole
set of samples globally. An influential example of global rules is the
Pseudoinverse rule\cite{Hagan,Marcus} used for training single-layer
networks.

One of the present authors has recently proposed another global learning
rule, called the Monte Carlo adaptation algorithm (shorten as MCA algorithm
hereafter)\cite{Zhao}. The basic idea is to make an adaptation to a randomly
chosen synaptic weight, and accept the adaptation if it improves the network
performance globally. A realization of the MCA algorithm had been used to
train single-layer feedback neural networks with binary-state synaptic
weights\cite{Zhao,Jin}.

The purpose of this Letter is to present a general version of the MCA
algorithm which is applicable to train multilayer neural networks with
either continuous or discrete synaptic states, and with either
differentiable or nondifferentiable neural transfer functions. Based on the
observation that the network performance is determined by the local fields
of neurons induced by the input of the samples (shorten as LFNIIS), our
algorithm is focused on controlling the distributions of LFNIIS by
continuously adapting the synaptic weights. Two steps are applied to perform
the control. The first one is to determine the target distribution of the
LFNIIS, define the states of synaptic weights, and chose the transfer
function of neurons for each layer respectively. The second one is to
randomly select a synaptic weight, and randomly adapt it to a new state,
then make a decision whether or not to accept this adaptation by a
criterion. This step is repeated till the distributions overlap with the
target ones. The criterion for acceptable adaptations is crucial for the
algorithm. In principle, we accept an adaptation if the distributions of
LFNIIS induced by it does not diverge away from the target distributions
statistically. This guarantees the distributions of LFNIIS evolving towards
the targets in a one-way manner.

As a realization example of the above framework, we train a two-layer neural
network to separate a set of linearly inseparable samples into two
categories. In order to demonstrate the network not only overcomes the
limitations of BP networks but also improves the network performance, we
emphasize its powerful generalization capacity over a two-layer BP network.
The generalization capacity is essential for a neural network. This is
because the sample set is normally representative of a much larger class of
patterns. It is particularly important that the network successfully
generalize what it has learned to the total population\cite
{Hagan,Stariolo,Sompolinsky}. In our example, when we input an unlearned
pattern having higher similarity with one sample, it is naturally desirable
that the network can classify it into the same category that the sample
belongs to. The degree of the average divergency of the pattern from the
sample below which the network can correctly categorize measures the
generalization capacity.

Suppose there are $M$ samples available for the training, and the $\mu $th
sample is represented by an $N$-dimensional binary vector ${\bf \xi }^{\mu
}\equiv \{\xi _{i}^{\mu },i=1,...,N\}$ with $\xi _{i}^{\mu }=\pm 1$. It has
been proved \cite{Gardner} that the maximum amount of samples that can be
linearly separable is $M\leq 2N$ if the samples have no correlations, and a
single-layer neural network trained by the perceptron rule can fulfil this
task. To solve linearly inseparable classification problems, one has to
apply multilayer neural networks.

Let ${\bf J}^{(l)}$ represent the weight matrix and ${\bf v}^{(l-1)}$
represent the input vector of the $l$th layer of a multilayer neural
network. The output of the $l$th layer is determined by the equations: 
\begin{eqnarray}
h_{i}^{(l)} &=&\sum_{j=1}^{N^{(l-1)}}J_{ij}^{(l)}v_{j}^{(l-1)},  \label{1} \\
v_{i}^{(l)} &=&\sigma ^{(l)}(h_{i}^{(l)}),  \label{2}
\end{eqnarray}
where $h_{i}^{(l)}$ is the local field of the $i$th neuron in the $l$th
layer and $\sigma ^{(l)}$ is its transfer function. In the equation, $%
N^{(l)} $ represents the number of neurons in the $l$th layer with $%
N^{(0)}\equiv N$.

A two-layer neural network has a hidden layer and a output layer. When the $%
\mu $th sample ${\bf \xi }^{\mu }$ is inputted into the network one obtains
the local field, $h_{i\mu }^{(1)}=\sum_{j=1}^{N}J_{ij}^{(1)}\xi _{j}^{\mu }$%
, and the output, $v_{i\mu }^{(1)}=\sigma ^{(1)}(h_{i\mu }^{(1)}) $, of the $%
i$th neuron in the hidden layer.

For the output layer, because we want the network to separate the samples
into only two categories, one neuron in this layer is enough. In this case,
the weight matrix will be a $1\times N^{(1)}$ matrix, whose elements will be
denoted by $J_{1j}^{(2)}$. Here $N^{(1)}$ is the number of neurons in the
hidden layer. Inputting the vector ${\bf v}_{\mu }^{(1)}$ to the output
layer one obtains the local field, $h_{\mu
}^{(2)}=\sum_{j=1}^{N^{(1)}}J_{1j}^{(2)}v_{j\mu }^{(1)}$, and the output, $%
v_{\mu }^{(2)}=\sigma ^{(2)}(h_{\mu }^{(2)})$, of the neuron.

Let $\Sigma _{1}$ and $\Sigma _{2}$ represent the two categories of samples.
Our goal is to establish the connections, $v_{\mu }^{(2)}=1\smallskip $ if $%
{\bf \xi }^{\mu }\in \Sigma _{1}$ and $v_{\mu }^{(2)}=-1\smallskip $ if $%
{\bf \xi }^{\mu }\in \Sigma _{2}$, by the proper solution of ${\bf J}^{(l)}$
and $\sigma ^{(l)}$. To fulfil this goal, the transfer function of the
neuron in the output layer must be the step function: $\sigma ^{(2)}(x)=1 $
for $x\geq 0$ and $\sigma ^{(2)}(x)=-1$ for $x<0$.

The establishment of the connections implies the satisfaction of the
condition $t_{\mu }h_{\mu }^{(2)}\geq 0$ in terms of the local fields, where 
$t_{\mu }=1$ for ${\bf \xi }^{\mu }\in \Sigma _{1}$ and $t_{\mu }=-1$ for $%
{\bf \xi }^{\mu }\in \Sigma _{2}$. However, this is not enough for the
generalization. When inputting a vector which has a set of elements, denoted
by $\{k\}$, different from, say, the $\mu $th sample, then the local fields
of the neurons in the hidden layer induced by this input should be $%
\overline{h}_{i\mu }^{(1)}=h_{i\mu }^{(1)}-2\sum J_{ij}^{(1)}\xi _{j}^{\mu }$%
, where the sum is over the set of $\{k\}$. This in turn results in a set of
elements, denoted by $\{k^{\prime }\}$, different from ${\bf v}_{\mu }^{(1)}$
and leads to $\overline{h}_{\mu }^{(2)}=h_{\mu }^{(2)}-2\sum
J_{1j}^{(2)}v_{j\mu }^{(1)}$ for the neuron in the output layer, where the
sum is over $\{k^{\prime }\}$. The generalization capacity is thus
determined by the capability of conserving the sign of $h_{i\mu }^{(1)}$ and 
$h_{\mu }^{(2)}$ under as many mutations as possible of the sample ${\bf \xi 
}^{\mu }$, which requires the absolute values of not only $h_{\mu }^{(2)} $
but also $h_{i\mu }^{(1)}$ as big as possible.

Thus, to gain better generalization capacity we should expect the
distribution of $h_{\mu }^{(2)}$ satisfy the condition $t_{\mu }h_{\mu
}^{(2)}\geq c$, where $c$ is a positive parameter. For the hidden layer,
there is no restriction on the sign of a specific $h_{i\mu }^{(1)}$, we thus
define $d_{i}=\sum_{\mu =1}^{\mu =M}|h_{i\mu }^{(1)}|$ to roughly measure
the mean absolute value of the local fields. To gain better generalization
capacity we expect $d_{i}$ to be as large as possible.

We apply the following procedure to train the network to find a set of
solution of synaptic weights that gurantees the desired distributions of $%
h_{i\mu }^{(1)}$ and $h_{\mu }^{(2)}$ be satisfied.

(1) Initialize $J_{ij}^{(l)}$ with $J_{ij}^{(l)}\in \{\theta
_{k}^{(l)},k=1,...,p\}$ randomly with equal probability; calculate $h_{i\mu
}^{(1)}$, $v_{i\mu }^{(1)}$, $h_{\mu }^{(2)}$ and $d_{i}$. Here $\theta
_{k}^{(l)}$ is a state of $J_{ij}^{(l)}$.

(2) Randomly select a $J_{ij}^{(l)}$ and randomly adapt it to a new state $%
\theta _{\widetilde{k}}^{(l)}$; if $l=1$ calculate 
\begin{eqnarray*}
\widetilde{h}_{i\mu }^{(1)} &=&h_{i\mu }^{(1)}+(\theta _{\widetilde{k}%
}^{(1)}-\theta _{k}^{(1)})\xi _{j}^{\mu }, \\
\widetilde{v}_{i\mu }^{(1)} &=&\sigma ^{(1)}(\widetilde{h}_{i\mu }^{(1)}), \\
\widetilde{h}_{\mu }^{(2)} &=&\sum_{q=1}^{N^{(1)}}J_{1q}^{(2)}\widetilde{v}%
_{q\mu }^{(1)}, \\
\widetilde{d}_{i} &=&\sum_{\mu =1}^{M}|\widetilde{h}_{i\mu }^{(1)}|;
\end{eqnarray*}
if $l=2$ calculate 
\[
\widetilde{h}_{\mu }^{(2)}=h_{\mu }^{(2)}+(\theta _{\widetilde{k}%
}^{(2)}-\theta _{k}^{(2)})v_{j\mu }^{(1)}\text{.} 
\]
Then calculate 
\[
n=\sum_{\{\mu \}}n_{\mu }, 
\]
where 
\[
n_{\mu }=\left\{ 
\begin{array}{c}
0,t_{\mu }(\widetilde{h}_{\mu }^{(2)}-h_{\mu }^{(2)})=0 \\ 
1,t_{\mu }(\widetilde{h}_{\mu }^{(2)}-h_{\mu }^{(2)})>0 \\ 
-1,t_{\mu }(\widetilde{h}_{\mu }^{(2)}-h_{\mu }^{(2)})<0
\end{array}
\right. , 
\]
and the sum is just{\it \ }over those index set $\{\mu \}$ of $\mu $
satisfying $t_{\mu }h_{\mu }^{(2)}<c$ or $t_{\mu }\widetilde{h}_{\mu
}^{(2)}<c$.

(3) If $\widetilde{d}_{i}\geq d_{i}$ and $n\geq 0$, renew the parameters,
i.e., $h_{i\mu }^{(1)}\longleftarrow \widetilde{h}_{i\mu }^{(1)}$ , $v_{i\mu
}^{(1)}\longleftarrow \widetilde{v}_{i\mu }^{(1)},$etc., otherwise remain
the old ones; return to step (2) till the condition $t_{\mu }h_{\mu
}^{(2)}\geq c$ is achieved and $d_{i}$ can not be further enlarged.

With a set of parameters $N=1000$, $N^{(1)}=1000$, $M=2400$, $c=30$, and
applying the binary weights $J_{ij}^{(l)}\in \{+1,-1\}$ while adopting the
step transfer function for each neuron, we tested the above training
procedure by separating the samples into two sets with equal samples.
Without loss of the generality, we suppose ${\bf \xi }^{\mu }\in \Sigma _{1}$
for $\mu =1,...,M/2$ and ${\bf \xi }^{\mu }\in \Sigma _{2}$ for $\mu
=M/2+1,...,M$. Note that the samples are linearly inseparable since $M>2N $.
In Fig. 1(a) and 1(b) the up-triangles show the distributions of $h_{i\mu
}^{(1)}$ and $h_{\mu }^{(2)}$ respectively. It can be seen that $h_{\mu
}^{(2)}$ distributes in the region of $t_{\mu }h_{\mu }^{(2)}\geq 30$
correctly, and the distribution of $h_{i\mu }^{(1)}$ shows a two-peak
structure.

The two-peak structure is a consequence of controlling the distribution of $%
h_{i\mu }^{(1)}$ by restricting $\widetilde{d}_{i}\geq d_{i}$ for acceptable
adaptations. If merely employ $n\geq 0$ as the criterion for acceptable
adaptations, the distribution of $h_{\mu }^{(2)}$ can fulfil the condition $%
t_{\mu }h_{\mu }^{(2)}\geq 30$ easily. However, the distribution of $h_{i\mu
}^{(1)}$ will be out of control. The open stars in Fig. 1(a) and 1(b) show
the distributions of $h_{i\mu }^{(1)}$ and $h_{\mu }^{(2)}$ respectively. It
can be seen that $h_{i\mu }^{(1)}$ distributes around the origin with a
single-peak structure, while the distribution of $h_{\mu }^{(2)}$ is similar
to that obtained by using $n\geq 0$ and $\widetilde{d}_{i}\geq d_{i}$ as the
criterion.

For the generalization, the distribution of $h_{i\mu }^{(1)}$ with the
two-peak structure is obviously preferable than that with the single-peak
structure, since the amount of elements of $h_{i\mu }^{(1)}$ with small
absolute values in the former case is much less than that in the latter
case. Figure 2 confirms this prediction. In the figure, the triangles and
stars show the generalization capacity of the networks obtained with the
criterion $\widetilde{d}_{i}\geq d_{i}$ and $n\geq 0$, and with merely the
criterion $n\geq 0$, respectively. The horizontal axis is the mean
percentage of the difference between an input vector and one of the samples.
The vertical axis is the rate of correct classification. It is clear that
the former network has much higher generalization capacity than the later
one. Note that an input vector with no correlation with any sample has equal
probability to be classified into either category, a rate of $0.5$ therefore
indicates the total loss of the generalization capacity.

By adopting the mean square error $<(h_{\mu }^{(2)}-c)^{2}>$ as the
performance index and the analog function $\sigma ^{(1)}(x)=tanh(x)$ as the
transfer function for each neuron in the hidden layer, one can obtain a
two-layer network capable of categorizing the same set of samples using the
BP algorithm. In order to make comparison with our network, we apply $c=34$
and stop the learning procedure after the condition $t_{\mu }h_{\mu
}^{(2)}\geq 30$ is satisfied for all samples. The weights are normalized to
satisfy $<J_{ij}^{(l)}>=1$. The dot-lines in Fig. 1 show the distributions
of the LFNIIS for the BP network. It can be found that $h_{\mu }^{(2)}$
distributes around $\pm c$ as two Guassian-like peaks, and $h_{i\mu }^{(1)}$
distributes around the origin. Clearly, the distribution of the LFNIIS for
the hidden layer is out of the control of the algorithm since $h_{i\mu
}^{(1)}$ is not included in the performance index, and the two peaks in the
distribution of $h_{\mu }^{(2)}$ is induced by the operation of minimizing
the mean square error. The minimization operation drives the local fields
not only with smaller value but also with larger value of $t_{\mu }h_{\mu
}^{(2)}$ concentrated towards $c$ synchronously, while the larger values are
favorable for the generalization capacity as explained earlier. Thus, the
generalization capacity of the BP network would be even worse than that of
our network obtained with merely the criterion $n\geq 0$. The rate of
correct generalization represented by the dots in Fig. 2 confirms this
prediction.

Our procedure can be directly extended to train neural networks with
synaptic weights having more discrete states. It can be found easily that
when the states are extended to infinite, e.g., $J_{ij}^{(l)}\in \{\pm 1,\pm
3,{\pm 5,...},\}$, the weights indeed become continuous (after been
normalized). We have observed that the network performance can be further
improved by increasing the discrete states. To show this, we made, between
the networks with $J_{ij}^{(1)}\in \{\pm 1\}$ and the networks with $%
J_{ij}^{(1)}\in \{\pm 1,\pm 3\}$, a comparison of the maximum capacity of
separating no-correlation samples into two sets with equal members. The
weights in the output layer are fixed at $J_{1i}^{(2)}\in \{\pm 1\}$ for
both networks. The results are shown in Fig.3, where $M/N$ is the normalized
maximum amount of the samples that can be separated into two sets correctly
within $0.01MNN^{(1)}$ times of repeat of the steps (2)-(3), and $N^{(1)}/N$
is the normalized number of neurons in the hidden layer. The up- and
down-triangles represent the results for networks with $J_{ij}^{(1)}\in
\{\pm 1\}$ and with $J_{ij}^{(1)}\in \{\pm 1,\pm 3\}$ respectively. In the
calculation we fix $N=500$. One can see from the figure that the maximum
capacity increases as the increase of the neurons in the hidden layer, and
increases with the increase of the discrete states of weights.

In summary, unlike the BP algorithm, the improved MCA algorithm puts no
restriction to the neural transfer function and is applicable to train
neural networks with either discrete or continuous synaptic weights. Another
key difference is that we implement the desired network performance by
controlling the distributions of the LFNIIS, while the BP algorithm
approaches this goal by minimizing the performance index defined constantly
as the mean square error. It is obvious that one has a much wider freedom to
improve the network performance by controlling the distributions of the
LFNIIS. This is because one has freedom not only to control the distribution
for the output layer but also to control the distributions for the hidden
layers. The good generalization capacity of the two-layer network trained
with the criterion $\widetilde{d}_{i}\geq d_{i}$ and $n\geq 0$ is just
benefited from the control of the distribution of the LFNIIS for the hidden
layer.

The application of the algorithm described in this Letter to the problem of
separating a set of samples into several categories is straightforward by
involving more neurons in the output layer. The algorithm is directly
applicable to train single-layer networks, and can be extended
straightforwardly to train networks with three or more layers.

We want to emphasize that the MCA algorithm has capability for practical
applications. For example, it takes about one hour of evolution time for a
personal computer to train the network satisfy the condition $t_{\mu }h_{\mu
}^{(2)}\geq 30$ by applying the criterion $n\geq 0$. To fulfil the same
condition using the BP algorithm with optimal learning rate it takes about 6
computer hours.

It is necessary to point out that the training procedure is sensitive to
technical details. For example, if one replaces the criterion $n\geq 0$
simply with $n>0$ in the related training preocedures above, it may need
double the time to approach the same goals. On the other hand, certain
treatments, such as adjusting the constant $c$ gradually to its target
value, can dramatically decrease the training time. In addition, introducing
temperature to the criterion for acceptable adaptations can affect the
efficiency of training process in a complex way. These facts imply that
there is a big possibility to further improve the training procedure.

Finally we briefly report an interesting phenomenon which may share lights
on the role of different layers in a network. We have performed the MCA
algorithm in two ways. One is to fix the wights in the output layer by some
random realizations and merely adjust the weights in the hidden layer. The
another is to adjust the weights in both layers. It was found that both ways
can achieve the same goal of classification, but the training time used in
the first way was dramatically less than that used in the second way. This
implies that the role of the output layer is merely to span out the space.
Each specific realization of weights for the output layer has a set of
optimal realizations of the weights for the hidden layer, and every
realization leads to the same target distributions of $h_{i\mu }^{(1)}$ and
(b) $h_{\mu }^{(2)}$.

Particular thanks are given to Professor Schuster from whom I got a lot of
useful ideas and suggestions related to this work. This work is supported in
part by the National Natural Science Foundation of China under Grant No.
10475067, and the Doctor Education Fund of the Educational Department of
China.

FIGURE CAPTIONS \newline
\ \ \ Fig.1 The distributions of LFNIS of (a) $h_{i\mu }^{(1)}$ and (b) $%
h_{\mu }^{(2)}$.

Fig.2. The generalization capacity of the networks.

Fig.3. The maximum capacity of classification as functions of the neuron
number in the hidden layer for $J_{ij}^{(1)}\in \{\pm 1\}$ and for $%
J_{ij}^{(1)}\in \{\pm 3,\pm 1\}$.


\begin{references}
\bibitem{Hagan}  M. Hagan, H. Demuth, and M. Beale, {\it Neural Network
Design,} Boston, MA: PWS Publishing, 1996.

\bibitem{Rumelhart}  D. E. Rumelhart and J. L. McClelland, eds., Parallel
Distributed Processing: Exlporations in the Microstructure of Cognition,
Vol. 1, Cambridge, MA: MIT Press, 1986.

\bibitem{Petersen}  C. C. H. Petersen, R. C. Malenka, R. A. Nicoll, and J. J.

Hopfield, Proc. Natl. Acad. Sci. U.S.A. 95, 4732(1998).

\bibitem{Connor}  D. H. O'Connor, G. M. Wittenberg, and S. S.-H. Wang, Proc.

Natl. Acad. Sci. U.S.A. 102, 9679(2005); G. M. Wittenberg and S. S.-H. Wang,
J. Neurosci. 26, 6610(2006).

\bibitem{Abarbanel}  H. D. I. Abarbanel, S. S. Talathi, L. Gibb, and M. I.
Rabinovich, Phys. Rev. E 72, 031914(2005).

\bibitem{Marcus}  C. M. Marcus, F. R. Waugh, and R. M. Westervelt, Phys.
Rev. A 41, 3355(1990)

\bibitem{Zhao}  H. Zhao, Phys. Rev. E 70, 066137(2004).

\bibitem{Jin}  T. Jin and H. zhao, Phys. Rev. E 72, 086512(2005).

\bibitem{Stariolo}  D. A. Stariolo and F. A. Tamarit, Phys. Rev. A 46,
5249(1992)

\bibitem{Sompolinsky}  H. Sompolinsky and N. Tishby, Phys. Rev. Lett. 65,
1683(1990)

\bibitem{Gardner}  E. Gardner, J. Phys. A: Meth. Gen. 21, 257(1988).
\end{references}
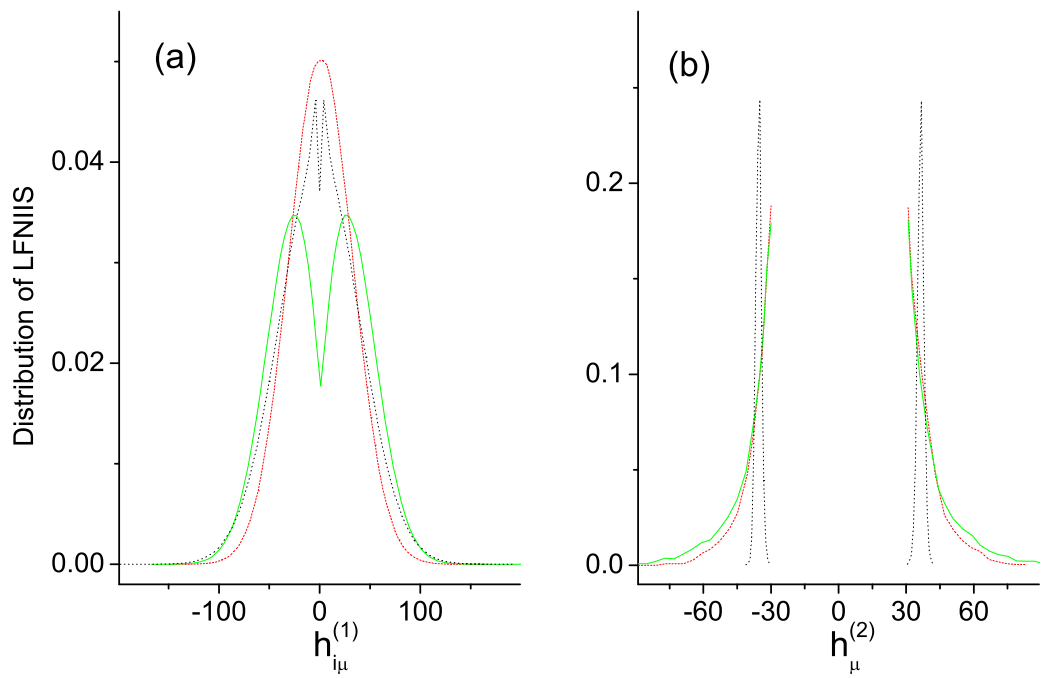
\end{document}